# Development and test of a real-size MRPC for CBM-TOF

Yi Wang[a], Pengfei Lyu[a], Xinjie Huang[a], Dong Han[a], Bo Xie[a], Yuanjing Li[a], Norbert Herrmann[b], Ingo Deppner[b], Christian Simon[b], Pierre-Alain Loizeau[b], Philipp Weidenkaff[b], Frühau Jochen[c], M.Laden Kis[c]

[a] *Key Laboratory of Particle and Radiation Imaging, Department of Engineering Physics, Tsinghua University, Beijing 100084, China*

[b] *Physikalisches Institut, University Heidelberg, Heidelberg, Germany*

[c] *GSI Helmholtzzentrum für Schwerionenforschung, GSI, Damstadt, Germany*

ABSTRACT: In the CBM (Compressed Baryonic Matter) experiment constructed at the Facility for Anti-proton and Ion Research (Fair) at GSI, Darmstadt, Germany, MRPC(Multi-gap Resistive Plate Chamber) is adopted to construct the large TOF (Time-of-Flight) system to achieve an unprecedented precision of hadron identification, benefiting from its good time resolution, relatively high efficiency and low building price. According to the particle flux rate distribution, the whole CBM-TOF wall is divided into four rate regions named Region D, C, B and A (from inner to outer). Aiming at the Region C and B where the rate ranges from 3.5 to 8.0 kHz/cm$^2$, we've developed a kind of double-ended readout strip MRPC. It uses low resistive glass to keep good performance of time resolution under high-rate condition. The differential double stack structure of 2x4 gas gaps help to reduce the required high voltage to half. There are 24 strips on one counter, and each is 270mm long, 7mm wide and the interval is 3mm. Ground is placed onto the MRPC's electrode and feed through is carefully designed to match the 100Ω impedance of PADI electronics. The prototype of this strip MRPC has been tested with cosmic ray, a 98% efficiency and 60ps time resolution is gotten. In order to further examine the performance of the detector working under higher particle flux rate, the prototype has been tested in the 2014 October GSI beam time and 2015 February CERN beam time. In both beam times a relatively high rate of 1 kHz/cm$^2$ was obtained. The calibration is done with CBM ROOT. A couple of corrections has been considered in the calibration and analysis process (including time-walk correction, gain correction, strip alignment correction and velocity correction) to access actual counter performances such as efficiency and time resolution. An efficiency of 97% and time resolution of 48ps are obtained. All these results show that the real-size prototype is fully capable of the requirement of the CBM-TOF, and new designs such as self-sealing are modified into the strip counter prototype to obtain even better performance.

# Contents

## 1. Introduction

Since its birth in 1996, large amount of efforts has been devoted into R&D of Multi-gap Resistive Plate Chambers (MRPCs) for improved performance including timing resolution and detecting efficiency, and large-area time-of-flight (TOF) system applied inexpensive MRPCs becomes widespread in modern nuclear and particle physics experiments in these years. In the CBM (Compressed Baryonic Matter) experiment constructed at the Facility for Anti-proton and Ion Research (Fair) at GSI, Darmstadt, Germany, MRPCs are also the best solution to construct the large TOF system for an unprecedented precision of hadron identification. The challenge for CBM is an extremely high requirement of MRPCs' rating capability, for the CBM-TOF will work at very high rate (~25kHz/cm$^2$) environment. At the same time, CBM-TOF is designed to work at free-running mode, thus the detectors' noise level must be very low. According to the particle flux rate distribution, the whole CBM-TOF wall is divided into four rate regions named Region D, C, B and A (from inner to outer). Aiming at the Region C and B where the rate ranges from 3.5 to 8.0 kHz/cm$^2$, we've developed a kind of double-ended readout Strip-MRPC. The prototype has been produced and tested in two beam tests in GSI and SPS CERN.

This article is organized as follows: Section 2 presents a general introduction of Strip-MRPC's structure; Section 3 is devoted to the performance of Strip-MRPC in GSI 2014 and CERN SPS 2015 beam time, including the analysis process and results; Section 4 describes a new slewing method of time-walk correction modified into analysis code and the results we obtain through it; Section 5 summarizes our conclusions and outlook.

## 2. Strip-MRPC module

This Strip-MRPC module is of differential double stack structure with 2x4 gas gaps, which can help to reduce the required high voltage to half. There are 24 strips on one counter, and each is 270mm long, 7mm wide and the interval is 3mm. Low resistive glass instead of float glass is applied to keep good performance of time resolution under high-rate condition. Ground is placed onto the MRPC's electrode and feed through is carefully designed to match the 100Ω impedance of PADI electronics, just to minimize the noise caused by reflection.

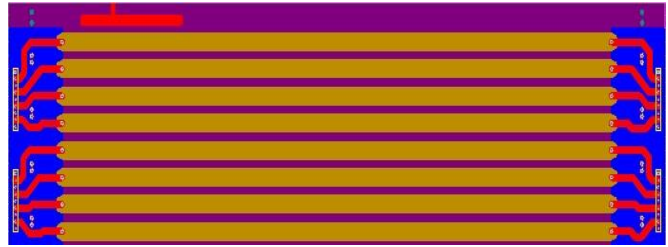

Figure 2.1 Strip-MRPC's readout PCB: strip and transmission line is designed to fit 50Ω impedance, and a differential combination of two readout PCB match the 100Ω impedance of PADI electronics.

## 3. Analysis results of beam test

Before the experimental setup of MRPC in two beam time is presented, how the CBM-Root do analysis on raw data should be explained first. The relative code, developed by CBM-TOF group, considers a series of influences factors and calibrate them out from the measured data: Time-walk correction corrects the dependence of measured time on strength of the analogue signal; Gain correction eliminate the amplification gain jitter of different PADI channel; Strip alignment removes the shifting center of different strip lead by cable length; Velocity correction calibrate the time difference brought by different particle speed. All these 4 corrections are looped over with different parameters by multiple times in the analysis until observables from fully calibrated hits include efficiency, time resolution and cluster size don't improve anymore.

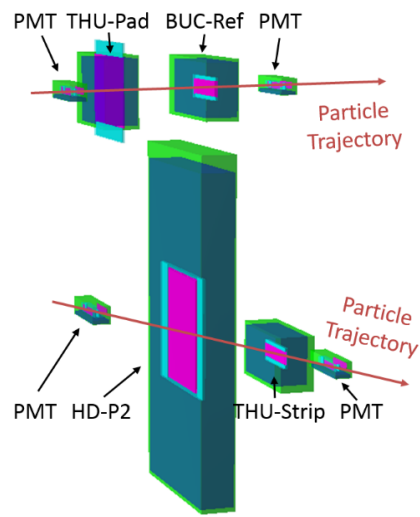

Figure 3.1.1 Beam time setup of GSI Oct 2014. THU-Pad and BUC-Ref are in upper setup, while HD-P2 and THU-Strip are among lower setup. PMTs are applied for purpose of counting rate calibration.

## 3.1 GSI Oct 2014 beam time

The experimental setup of this beam time is shown in figure 3.1.1, which is generated by CBM Root geometry file. All of these tested MRPC modules are divided into two parts, and the Strip-MRPC (THU-Strip in figure) module is among the lower setup. In this beam test, we have 1.1A GeV $^{152}$Sm beam on 0.3mm/4mm/5mm Pb target, and a flux rate of several hundred Hz/cm$^2$ is available.

In order to get the Strip-MRPC's performance, we define it the detector under test (Dut) in the analysis. HD-P2 counter is Mref, and only with a reference counter all the information (time difference, time resolution, efficiency, etc.) we want for Dut can be obtained. We need another counter to give the starting time of one event required by velocity correction, thus Diamond counter, which is not shown in the figure above, is applied to work as so-called Bref. After setting cut parameters to kick out noise and do hits selection, all the calibration modes are carried out in an iterative way until all unrelated influence on time of flight is calibrated out. Then we get the calibrated time resolution of detector under test. For run Sun1205 with Strip-MRPC's HV 5500V and threshold 200mV, After projecting the time velocity correction histogram to the y scale that is the time difference between Dut and Mref, we get a system time resolution of 68.86ps shown in figure 3.1.2. Asuming the THU-Strip and HD-P2's time resolution are the same, an independent timing ablity of $68.86/\sqrt{2} = 48.7\, ps$ is calculated.

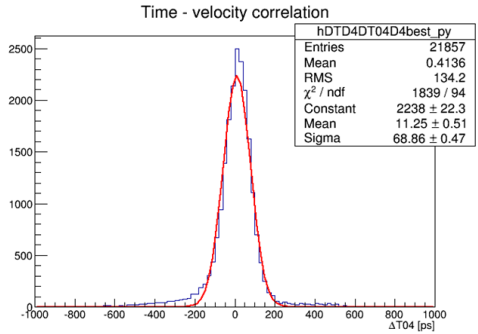

Figure 3.1.2 System time resolution of THU-Strip and HD-P2 in run Sun1205 achieves 68.9ps in GSI Oct 2014 beam time.

All four available runs with different PADI threshold from 170mV to 200mV have been analyzed. As is shown in figure 3.1.3, the time resolution is around 50ps, efficiency 97% and cluster size 1.6. All of these results are very stable for change in threshold is not significant.

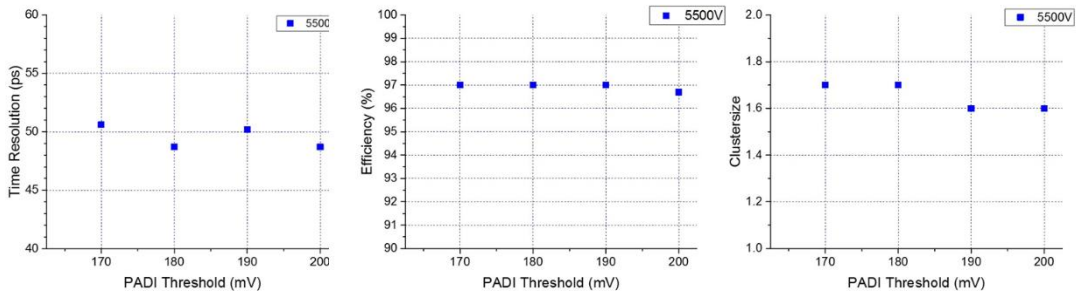

Figure 3.1.3 Time resolution (around 50ps), efficiency (97%), clustersize (1.6 to 1.7) of Strip-MRPC under different PADI electronics threshold.

## 3.2 SPS Feb 2015 beam time

For SPS beam time in Feb 2015, the setup is also divided by two parts, and more counters are added in, which means more combinations of these counters to be Dut, Mref and Bref. This time 13A GeV Ar beam is applied, and the rate has thus increased to 1kHz/cm$^2$.

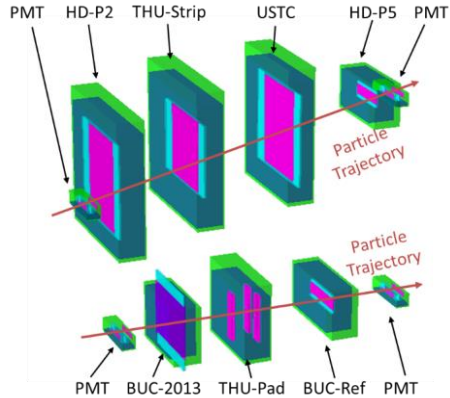

Figure 3.2.1 Beam time setup of SPS Feb 2015. HD-P2, THU-Strip, USTC and HD-P5 are in upper setup, while BUC-2013, THU-Pad, BUC-Ref are among lower setup. PMTs are applied for purpose of counting rate calibration.

Diamond counter is not available in this test, so we need to get counters for MRef and Bref among the 3 detectors except THU-Strip in the upper setup. Among all the cominations, we've got the best result of Dut THU-Strip by setting HD-P2 Mref and USTC Bref. The analysis sequence is same to that of Oct's analysis. To get better calibration result, some cut selection is done, especially on the Y position selection along the strip and chi2 selection limit on the time difference distribution between THU-Strip and Mref. When the cut is set to 10 of chi2 limit and 0.3 of Y Position selection, we can get a relatively best result and for run01Mar1126, the system time resolution is 96ps.

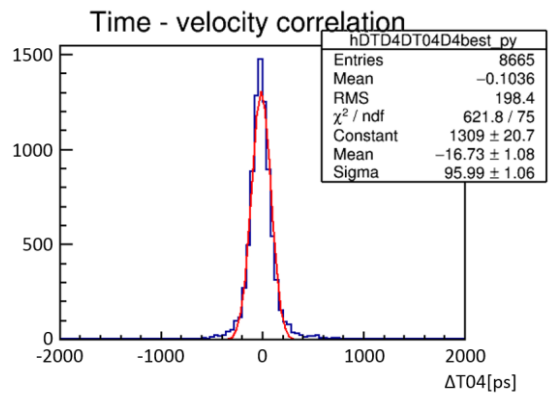

Figure 3.2.2 System time resolution of THU-Strip and HD-P2 in run01Mar1126 achieves96.0ps in Sps Nov 2015 beam time.

Apparently, the time resolution in this beam time is not as good as in GSI Oct 2014. One possible reason is diamond detector is not applied as Bref anymore, thus the distance between MRef and BRef is closer leading that the velocity correction is not calibrated in an ideal way. Lately the tacking method is introduced by TOF group. Tracks are built in the four counters to cut efficiency and correct time and space for hits in Dut. After tracking, time resolution is greatly improved by 10 to 20 ps. Since the rate information is recorded throughout this beam time, we choose several runs at different flux rates and get the results shown in figure 3.2.3. The efficiency is stable with growth of rate, beacause low-resistive glass Strip-MRPC apply can maintain its performance under high rate. From the time resolution, we can easily find out the tracking method has done a lot benefit.

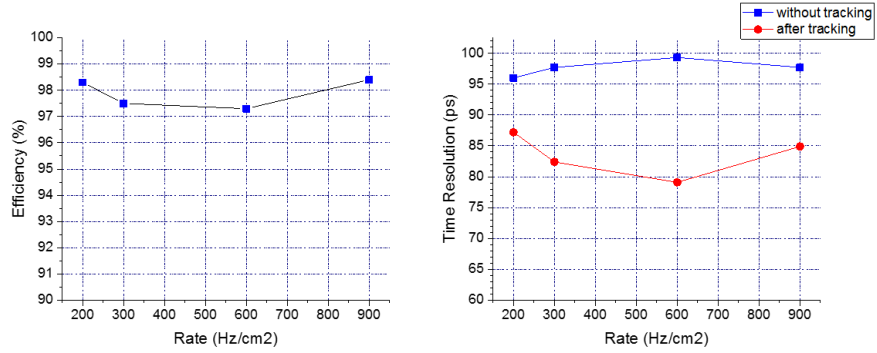

Figure 3.2.3 Efficiency (98%) and system time resolution of Strip-MRPC under different flux rate. The efficiency under rate 300Hz/cm$^2$ and 600Hz/cm$^2$ is relatively low because these two runs have less events. From the time resolution, tracking has provided an improvement up to 20ps.

## 4. Slewing correction

We also introduce a new correction method called slewing correction to the time-walk correction process in CBM Root. As mentioned before, the time difference is calibrated to be independent from the time over threshold (TOT), thus the mean value of each bin in the 2D plot should be a horizontal line. The original method takes the subtraction of mean time of each bin and average time of all bins as the correction, and for points between centres of each bin, the correction value is got by linear interpolation. While inn STAR, RHIC, an equation is introduced to fit the mean value of each bin, and the subtraction of this fitting value of each point and the total mean value will be the correction.

$$y = par[0] + par[1]/\sqrt{x} + par[2]/x + par[3] \times x$$

The equation has 4 parameters, and we modify the code of slewing methods to relevant CBM Root macros. These two plots show the independence level of time difference and TOT right after the step of walk correction. It can be noticed that in the red circle, the mean value line is obviously flatter in slewing method.

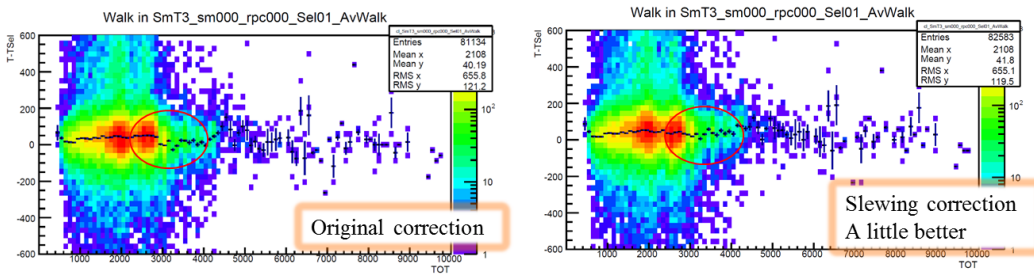

Figure 4.1 Walk correction results of original method and slewing correction method. In the slewing correction the mean value line of time difference is a little better.

We also scan the time resolution under different condition in both two beam time. The time resolution is always 1ps better, and this is outside the error bar. This modification has done some good to the walk correction.

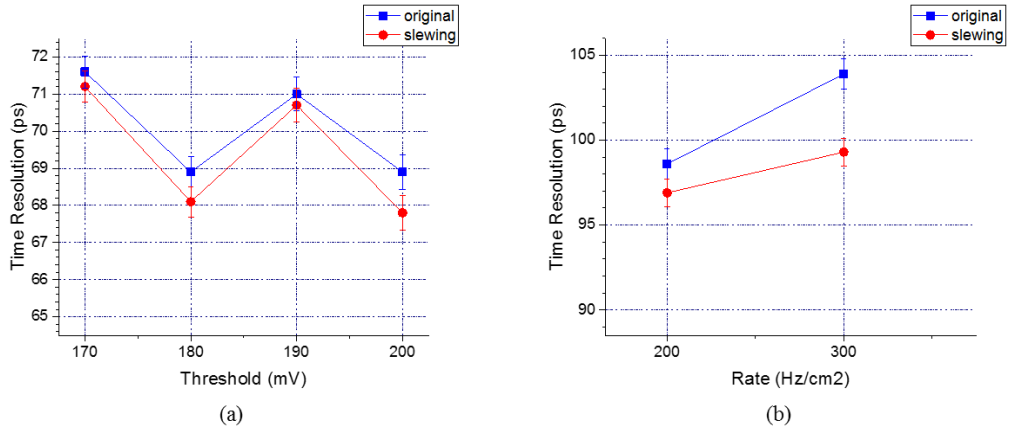

Figure 4.2 Time resolution difference between original walk correction and slewing method in GSI Oct 2014 (a) and SPS Feb 2015 (b), and slewing method always owns a better calibrated result.

## 5. Conclusion

The Strip-MRPC produced by Tsinghua University is proved to have an efficiency above 97%, time resolution below 60ps and cluster size around 1.6 through the two beam time of GSI Oct 2014 and SPS Feb 2015. Equipped with low-resistive glass, Strip-MRPC can maintain these good performances under high flux rate up to 10 kHz/cm$^2$. This Strip-MRPC is fully capable of meeting the demands of TOF wall. Some attempts on walk correction of CBM Root macros have been carried out and some progress is made. Another 3 strip prototypes with similar structure and real dimension have been developed and tested at SPS in Nov 2015, and the data analysis is undergoing.

## Acknowledgments

The work is supported by National Natural Science Foundation of China under Grant No.11420101004, 11461141011, 11275108. This work is also supported by the Ministry of Science and Technology under Grant No. 2015CB856905.